\RequirePackage{fix-cm}
\documentclass[smallextended]{svjour3}       
\smartqed  
\usepackage{graphicx}
\usepackage{color}
\usepackage{ulem}
\newcommand{\Eq}[1]{Eq. \ref{#1}}
\newcommand{\Eqs}[2]{Eqs. \ref{#1} and \ref{#2}}
\newcommand{\source}[4]{{\it #1}~{\bf #2}, #3 (#4)}
\newcommand{\ket}[1]{| #1 \rangle}
\newcommand{\bra}[1]{\langle #1 |}
\newcommand{\be}{\begin{equation}}
\newcommand{\ee}{\end{equation}}
\newcommand{\bea}{\begin{eqnarray}}
\newcommand{\eea}{\end{eqnarray}}

\newcommand{\eg}{{\it eg.}}

\newcommand{\etal}{{\it et. al.}}

\newcommand{\Sch}{Schr{\"o}dinger}
\newcommand{\oort}{\frac{1}{\sqrt{3}}}
\newcommand{\oord}{\frac{1}{\sqrt{d}}}

\newcommand{\third}{\frac{1}{3}}
\newcommand{\om}{\omega}

\newcommand{\CP}{{\cal P}}
\newcommand{\CHP}{{\cal H}_p}
\newcommand{\Zsdag}{Z_s^{\dag}}

\begin{document}

\title{Pointers for Quantum Measurement Theory}



\author{Jay Lawrence}

\institute{Department of Physics and Astronomy, Dartmouth
          College, Hanover, NH 03755, USA   \\
           The James Franck Institute, University of Chicago, 
           Chicago, IL 60637  \\ 
            \email{jay.lawrence@dartmouth.edu}  }         
         
\date{Received: date / Accepted: date}

\maketitle

\begin{abstract}
In the iconic measurements of atomic spin-1/2 or photon polarization, one employs 
two separate noninteracting detectors.  
Each detector is binary, registering the presence or absence of the atom or the 
photon.  For measurements on a $d$-state particle, we recast the standard von 
Neumann measurement formalism by replacing the familiar pointer variable with an 
array of such detectors, one for each of the $d$ possible outcomes.   We show that 
the unitary dynamics of the pre-measurement process restricts the detector outputs 
to the subspace of single outcomes, so that the pointer emerges from the apparatus.  
We propose a physical extension of this apparatus which replaces each detector 
with an ancilla qubit coupled to a readout device. This explicitly separates the pointer 
into distinct quantum and (effectively) classical parts, and delays the quantum to 
classical transition.  As a result, one not only recovers the collapse scenario of an 
ordinary apparatus, but one can also observe a superposition of the quantum 
pointer states.

\end{abstract}

\section{INTRODUCTION}
\label{intro}
The concept of a pointer was implicit in the quantum measurement formalism (1932) 
of von Neumann \cite{VN.55}.  It was developed later and used in parallel with 
decoherence theory starting in the 1970s.  Perhaps the most widely quoted 
definition was given in a recent review (2015) by Brasil and de Castro (BdC) 
\cite{Brasil}: ``Pointer states are eigenstates of the observable of the measurement 
apparatus that represent the possible positions of the display pointer of the equipment.'' 

Von Neumann's formalism begins with a separable state of the compound system
consisting of the object being measured (a spin, say), an apparatus, and an 
observer.  The spin interacts with the apparatus and entangles with it, and 
then the apparatus interacts with the observer and entangles with her.  If the
evolution up to this point is unitary, then it leads to the macroscopic superposition
in which, if the object is in the state $k$ ($k = 0,1,...,d-1$), then the ``display pointer 
of the equipment'' reads $k$, and the observer perceives the reading $k$.  However 
we, as observers, are unaware of the superposition; we perceive just a single 
outcome, as if chosen at random.  Von Neumann raised the question at what point 
the state vector collapses, including the possibility of the consciousness of the 
observer.  While he remained noncommittal on the answer, certain others 
believed that the site of collapse is indeed the consciousness of the observer 
\cite{London.Bauer.39,Wigner.61}.

Wigner \cite{Wigner.61} presented a scenario in which his ``friend'' is inserted 
into the sequence between the apparatus and Wigner himself.  Wigner then 
(hypothetically) interrogates and finds that his friend perceives only a single 
outcome.  He concludes that collapse occurred in the consciousness of his friend.

A different conclusion following from the same scenario 
was reached by Everett in 1957 \cite{Everett}.  Everett argued that the state vector 
need not collapse, because the state of an observer's consciousness, on any 
particular branch of the state vector, registers a single unambiguous outcome, 
leaving her blind to the existence of alternate outcomes expressed on other 
branches.  So she perceives collapse, even though the state vector (which 
includes her as a subsystem) does not collapse.  This is Everett's relative state 
picture. The appellation ``Many Worlds'' was provided later by De Witt and Graham 
\cite{DeWitt.Graham.73}.  

While the foregoing comments provide background and motivation, 
this paper will not be concerned further with, nor will it make use of any 
particular interpretation of quantum theory.   We refer interested readers to 
excellent accounts of interpretations and their historical context found
elsewhere.\footnote{See the book by Max Jammer \cite{Jammer}, the review 
article by de Castro et. al. \cite{de Castro} (especially pp. 30-35 for paradoxes 
and entanglement), and the textbook by Weinberg \cite{Weinberg}, 
which includes discussion (pp. 81-95) of decoherence together 
with interpretations.} Here we shall use only the generally accepted 
quantum theory as found in standard textbooks. 

Everett's theory paved the way for decoherence theory,
which, in its original form, is also interpretation-independent.  Its seminal 
contributions were nevertheless inspired (or at least informed) by Everett's 
work.  In 1970, Dieter Zeh \cite{Zeh.70} coined the term ``pointer,'' and argued 
that, as a macroscopic system, it is incapable of exhibiting perceptible 
superpositions of pointer positions.  He thus identified the crucial point that the 
invisibility of such superpositions allows for the internal consistency of unitary 
quantum theory.  In 1981-82, Zurek \cite{Zurek.81,Zurek.82} addressed the role 
of the environment specifically in resolving the so-called {\it preferred basis} 
problem, arguing that interactions between the apparatus and the environment 
determine the pointer basis of the apparatus.  The blindness of the observer to 
alternate outcomes arises formally here from the trace over unobserved states 
of the environment, which produces a (non-unitary) transformation of the 
object/apparatus system to a mixed state.   
While decoherence theory thus accounts for the ``apparent'' 
collapse, Zurek leaves open the question of ``whether, where, when, or how 
the ultimate collapse occurs,'' implicitly rejecting the many worlds view. 
 
A general theory of decoherence was brought forth in 1985 by Joos and Zeh 
\cite{JoosZeh.85}, covering a broad range of phenomena beyond the controlled
measurements envisioned by von Neumann, and ranging from the stability of 
chiral molecules to the motion of a dust grain in the atmosphere. This work, 
together with Zurek's, provided
a formal background for many calculations of decoherence effects as well as 
conceptual developments in the years to follow.  Comprehensive reviews were
published in the first few years of this century \cite{Zurek.03,Schloss.04,Joos.03}.
A more recent comprehensive overview was presented in 2014 by Schlosshauer
\cite{Schloss.14}, with selected applications to experimental studies and the
mitigation of decoherence effects in quantum information applications.  In a 
few concluding paragraphs he offers general comments on foundational
implications of the orthodox theory.\footnote{Orthodox refers in this context to the 
use of unitary evolution up to the point of the trace over environmental states, 
without the introduction of a phenomenological nonlinear interaction which 
would induce a collapse of the state vector.}  Briefly, it is  viewed as a
practical, interpretation-independent body of results obtained using standard 
quantum theory.  It does not solve the measurement problem because the trace 
over environmental states and the interpretation of the density matrix assume
the collapse postulate and the Born rule.  This conclusion might seem  ironic 
given the original impetus for the theory \cite{Everett,Zeh.70}, but it is reflected in 
the diverse range of interpretational views (if any) expressed by its practitioners.  

Of particular relevance for us is the insightful 2015 review of pointer states by Brasil 
and deCastro \cite{Brasil}.   While focusing on Zurek's approach, I believe that it 
accurately represents the current general understanding of the concept and usage 
of the pointer.  Beyond this, the review suggests a new and more explicit definition 
(quoted in our first paragraph) alluding to the dual nature of the pointer, where 
``eigenstates of the observable of the apparatus'' are distinguished from the states 
of the ``display pointer of the equipment.'' The distinction is not, however, made 
explicit in the mathematical representation of pointer states in the measurement 
chain \cite{Brasil,Zurek.81,Schloss.04}.  The 
same set of states (implicitly the same physical system) is assumed to entangle first 
with the object, at the premeasurement stage, and later with the environment, forming 
a mixed state of the object/pointer system and establishing a preferred basis for the 
pointer.  The only implied distinction is in the nature of the entanglement.  We argue 
that the distinction should go deeper, because the same physical system cannot 
play both entangling roles.
  
The ``display pointer,'' (in our case the detector array) is macroscopic and cannot 
be isolated from the environment - most essentially from its own internal degrees of 
freedom.  Therefore it cannot form a pure entangled state with the object as implied 
by the first link in the von Neumann chain \cite{Brasil,Zurek.81,Schloss.04}, together 
with the statement of basis ambiguity.\footnote{See Eq. 1 of Ref. \cite{Brasil}; Eqs. 1.1 
and 1.3 of Ref. \cite{Zurek.81}; Eq. 2.1 of Ref. \cite{Schloss.04}; and the statements 
of basis ambiguity following these equations.}  For measurements described by
this formalism, a separate physical system, within the apparatus but isolated from the 
environment, must play the role of entangling with the object - we will call this system 
the ``quantum pointer,'' with examples to be discussed shortly.  

The quantum pointer remains entangled with the object and isolated from the
environment until it is ``read'' by the display pointer.  This reading is an irreversible
decohering process which takes the spin/quantum pointer system from a pure
entangled state to a mixture.  The interaction between the quantum and display
parts of the pointer determines the ``preferred basis,'' which is the ``observable of 
the apparatus.''  Describing how the measurement process unfolds in this picture, 
including the roles of the quantum pointer, the display pointer (as an array of binary 
detectors), and the environment, is a major goal of this paper.  We will organize the 
discussion by answering specifically the following questions:

\bigskip
\centerline{{\bf Questions}}
\medskip

First, what is the quantum pointer?  That is, what physical system supports
the quantum ``pointer states,'' as distinct (in the words of BdC) from the ``display 
pointer of the equipment,'' which is effectively classical?  We present a physical 
realization of the quantum pointer as a set of qubits, separate from but coupled to 
readout devices which serve as the ``display pointer.''   We describe this system in 
the next section. 

The second question - What is ``the observable of the apparatus'' of which the
pointer states are eigenstates?   And a larger question - what is the complete set
of commuting observables needed to characterize the object/quantum-pointer 
system? In Section III we answer these questions, and we show how these 
observables can be measured experimentally.

Finally, how in detail is the preferred basis actually chosen?   We provide an 
unconventional answer in Sec. IV: It  involves the controlled interaction 
between the quantum and display parts of the pointer, as well as the 
decohering effect of the environment on the display part.

\section{The QUANTUM POINTER}
\label{sec:2}
To demonstrate the emergence of a quantum pointer, we begin with the model of 
a Stern-Gerlach measurement, and we trace the development of entanglement 
between a $d$-state object, pictured as a spin of arbitrary
dimension $d= 2S + 1$, and an array of $d$ qubits, serving as virtual detectors, 
whose states can be read out later with standard readout devices.  We follow
the steps of Ref. \cite{JL1}, which treated the $d=2$ case.  We choose the initial spin 
state to maximize the entanglement which will develop with the ancilla qubit 
system.  Illustrating with a spin-1 atom as a simplest nonbinary object, the 
desired initial state of the system at time $t_1$, prior to measurement, is
\be
   \ket{\Psi(t_1)} = \oort~\ket{\phi(\vec{r},t_1)} 
   \bigg( \ket{0}_s + \ket{1}_s + \ket{2}_s \bigg) \ket{000}_a,
\label{state1}
\ee
where the prefactor is the atom's spatial wave-function, and the post-factor is the initial 
state of the ancilla system, with all three of its qubits in their 0 states.  A magnetic field 
gradient then separates the atomic spin states $\ket{k}_s$ ($k = 0,1,2$) along different 
paths, shown schematically in Fig. 1,
\be
    \ket{\Psi(t_2)} = \oort \bigg( \ket{\phi_0(\vec{r},t_2)} \ket{0}_s + 
    \ket{\phi_1(\vec{r},t_2)} \ket{1}_s + \ket{\phi_2(\vec{r},t_2)} \ket{2}_s \bigg) \ket{000}_a,
\label{state2}
\ee
whose corresponding wave-packets $\ket{\phi_k(\vec{r},t_2)}$ are assumed to become 
orthogonal by the time $t_2$.   Following this, the atom interacts with the ancilla qubit 
$(a_i)$ on its path $(i)$.  The interaction is local and spin-independent; the presence 
of the atom at the site of $(a_i)$ induces its transition from 0 to 1, preserving the 
atom's spin.  The resulting entangled state at $t_3$ is
\be
    \ket{\Psi(t_3)} = \oort \bigg( \ket{\phi_0(\vec{r},t_3)} \ket{0}_s \ket{100}_a + 
    \ket{\phi_1(\vec{r},t_3)} \ket{1}_s \ket{010}_a + \ket{\phi_2(\vec{r},t_3)} \ket{2}_s 
    \ket{001}_a \bigg).
\label{state3}
\ee
Finally, we imagine bringing the paths back together at $t_4$, and if this can be done
coherently, we can write
\be
   \ket{\Psi(t_4)} = \oort ~\ket{\phi(\vec{r},t_4)} \bigg( \ket{0}_s \ket{100}_a + 
   \ket{1}_s \ket{010}_a +  \ket{2}_s \ket{001}_a \bigg),
\label{state4a}
\ee
showing entanglement between just the spin and the ancilla system.  We will return
to the issue of coherence in a later section.


\begin{figure}[h!]
\includegraphics[scale=0.47]{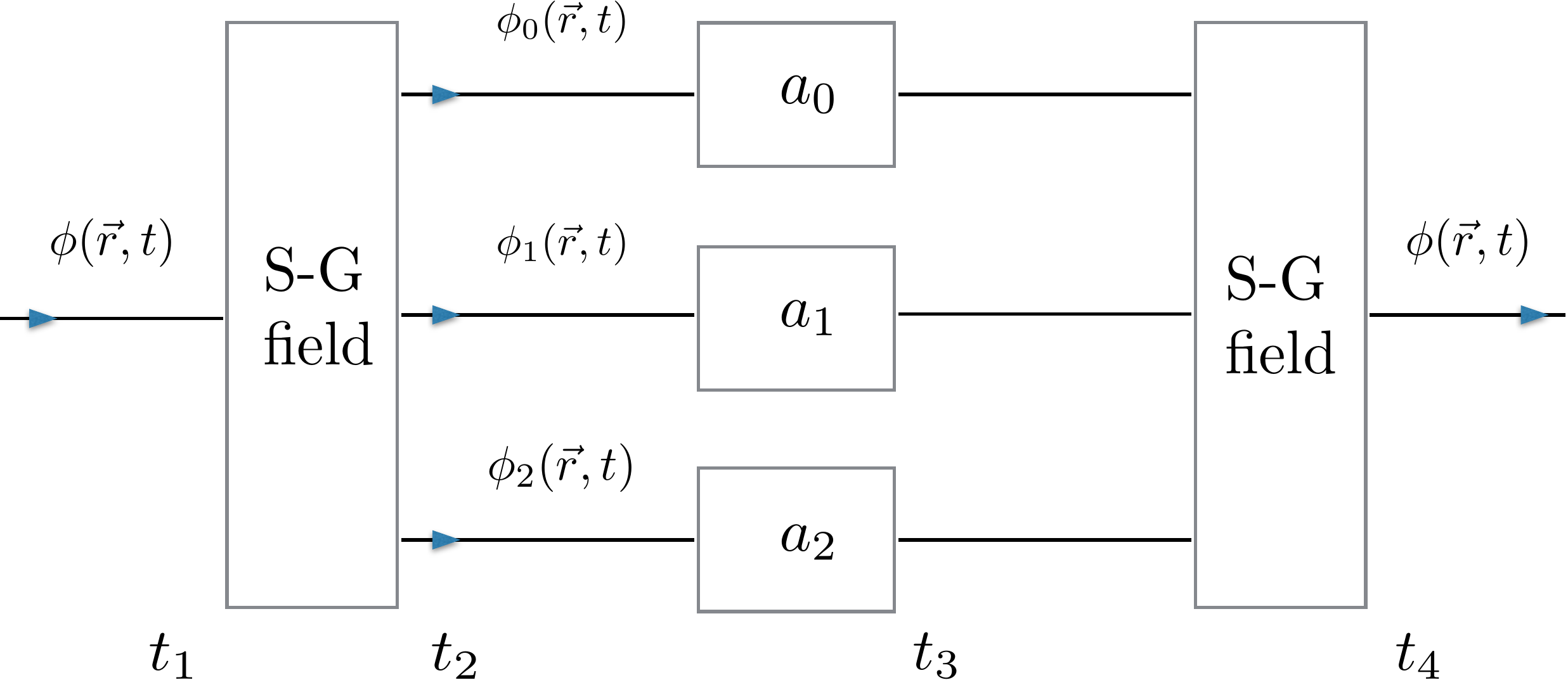}
\caption{\label{fig1} Stern-Gerlach schematic showing the evolution described by
Eqs. \ref{state1} - \ref{state4a}.}
\end{figure}

The process just described is unitary, and it comprises the premeasurement stage
of the full measurement process.  The resulting entangled  spin/ancilla system is 
the system of interest for us - we will describe measurements aimed at exploring 
its joint properties.  But first, as a preliminary, let us define the quantum pointer.

By the time $t_3$, the ancilla system has evolved  from its excitation vacuum state, 
$\ket{000}_a$ ($N=0$), to single-excitation states ($N=1$), as shown in Table I.  
Further operations, including all transformations and measurements to be performed
on the spin/ancilla system, will preserve the ancilla system in this $N=1$ subspace.  
This subspace defines the pointer:  The pointer is the effective qutrit whose Hilbert 
space $\CHP$ is spanned by the single-excitation states.  We will call these states 
$\ket{k}_p$ ($k = 0,1,2$, and $p$ for pointer), and in this notation, \Eq{state4a}  
(dropping the common spatial factor) becomes
\be
 \ket{\Psi} = \oort \bigg( \ket{0}_s \ket{0}_p + 
 \ket{1}_s \ket{1}_p + \ket{2}_s \ket{2}_p \bigg),
\label{state4b}
\ee
which indicates a generalized Bell state of the spin-pointer system.  Henceforth
$\ket{\Psi}$, without the temporal argument, is understood as the state of the 
spin/pointer system, as prepared at the time $t_4$.

Clearly this state generalizes to a spin of arbitrary dimension $d$, and a system of 
$d$ qubits (dimension $2^d$), whose pointer subspace shares the dimension $d$ 
with the spin.  Generalizing \Eqs{state4a}{state4b} (respectively) we have
\be
     \ket{\Psi} = \oord \sum_{k=0}^{d-1} \ket{k}_s ~x_{k} \ket{0,...,0}_a 
      = \oord \sum_{k=0}^{d-1} \ket{k}_s \ket{k}_p,
\label{state5}
\ee
where $x_{k}$ (a Pauli matrix) flips the spin of the $k$th qubit, generating the 
pointer state $\ket{k}_p$.  

\begin{table}
\caption{Basis states of the ancilla system with three qubits.  The pointer Hilbert 
space is spanned by the three singly-excited basis states shown.  Other states 
are dynamically excluded.}
\label{tab:1}       
\medskip
\begin{tabular}{|c|c|c|}
\hline
\ N \  & \ subspace bases \  & \ interpretation \   \\ \hline
 0 & \  $\ket{000}$ \ & \  initialized state \  \\ 
 1 & \  $\ket{100},~\ket{010},~\ket{001}$ \ & \ basis of pointer \  \\ 
 2 & \  $\ket{110},~\ket{101},~\ket{011}$ \  & \ excluded  \  \\ 
 3 & \   $\ket{111}$ \ & \ excluded \  \\  \hline
\end{tabular}
\end{table}

The restriction to the pointer subspace is enforced by the unitary dynamics of the 
premeasurement Stern-Gerlach evolution, and it gives rise to the {\it singleness} 
property - the fact that the measurement process will produce a single, 
unambiguous outcome for an observer or a recording device.  There is an 
observable, $N$, representing the total excitation number of the ancilla system.  
It is the sum of excitation numbers, $n_k = 0$ or 1, of the individual qubits $k$ 
($k = 0,..., d-1$), and it is given by
\be
   N= \sum_{k=0}^{d-1} n_k = \sum_{k=0}^{d-1} \big( 1 + z_k \big)/2,
\label{Nop}
\ee
where $z_k$ is the Pauli $z$-matrix (with eigenvalues $\pm 1$) acting on the 
$k$th qubit.  One can measure $N$ most simply by measuring every individual 
$z_k$ and applying (\ref{Nop}).  Acting on state (\ref{state5}), the measurement 
would show that $N = 1$, demonstrating the singleness property as listed in 
Table II.  This dramatic correlation within the ancilla system extends to the spin
qudit, whose readout (a state index $k$) corresponds to the output ($k$) of the 
display pointer.  This latter correlation is the projection property - the 
post-measurement spin state is the state identified by the display pointer.  
(This is called the ``post-measurement state-update rule.'')

We have just described the quantum pointer, as distinct from the display pointer.
It is ``quantum'' in two respects that the display pointer is not:  It is reversible,
and it can be prepared {\it and detected} in a superposition of its basis states.
That is, no preferred basis set is imposed on it.  We will show an example 
after introducing the relevant observables in the next section.

\begin{table}
\caption{Eigenvalues and Physical Interpretations}
\medskip
\begin{tabular}{|c|c|c|}
\hline
\  eigenvalue \  & \  property \  & \ physical interpretation  \   \\ \hline
$N = 1$ & \  singleness \ & \ one and only one detector will register  \  \\ 
$Z_s^{\dag}Z_p = 1$ & \  projection \  & \ post-measurement spin $\Leftrightarrow$ 
which detector \  \\ 
$X_sX_p = 1$ & \  superposition \ & \ $\ket{\psi(t_4)} =$ superposition of collapse scenarios \  \\ \hline
\end{tabular}
\smallskip
\end{table}

\section{OBSERVABLES  of the SPIN/POINTER SYSTEM}
\label{sec:3}
In the previous section we introduced observables for individual qubits in the
ancilla system - the usual Pauli matrices (denoted by lower case $z_k$ and 
$x_k$).  Here we require observables for two $d$-dimensional systems - the
spin and the quantum pointer.  These are the generalized Pauli operators
(representable as $d \times d$ matrices) which we denote by upper case:  
$Z_s$ and $X_s$ for the spin, and  $Z_p$ and $X_p$ for the 
quantum pointer.  The $Z$ are diagonal in the standard basis:
\be
   Z = \sum_{k=0}^{d-1} \ket{k}~\om^k \bra{k},
\label{bigZ}
\ee
where $\om = \exp(2 \pi i / d)$ is the $d$th root of unity.  Although $Z$ is not
hermitian, we refer to it as an observable because it is unitary, and its 
{\it exponents} $k$ are real numbers ($k = 0,1,...,d-1$), which label the 
eigenstates.  [In the case of spin, the $\hat{z}$ component of the physical 
spin takes the values $S_z =  -S+k$, where $S=(d-1)/2$.]  

The canonical conjugates of $Z$ are defined by
\be
   X = \sum_{k=0}^{d-1} \ket{k+1} \bra{k},
\label{bigX}
\ee
where the state index $k$ is defined modulo $d$.  The eigenbasis of $X$,
defined by
\be
     X \ket{k}_X = \omega^k \ket{k}_X,
\label{Xstates}
\ee
is the inverse quantum Fourier transform \cite{NC} of the standard basis; that is,
\be
  \ket{k}_X = {\cal F^{\dag}} \ket{k} =  \frac{1}{\sqrt{d}} \sum_{k'} \om^{-kk'} \ket{k'}, 
\nonumber
\ee
\be  
  \hbox{where}  \hskip0.6truecm
   {\cal F} \equiv \frac{1}{\sqrt{d}} \sum_k \sum_{k'} \ket{k'} ~\om^{kk'}\bra{k},
\label{Fourier}
\ee
and ${\cal F}$ and ${\cal F^{\dag}}$, being symmetric, are complex conjugates.

The correlations of interest are described by tensor product operators; the simplest
example is $\Zsdag Z_p$, which clearly has \Eq{state5} as an eigenstate with 
eigenvalue unity,
\be
   Z_s^{\dag} Z_p \ket{\Psi} = \ket{\Psi}.
\label{projection}
\ee
So, while measurement outcomes of $Z_s$ and $Z_p$ are separately random, they
are perfectly correlated - \Eq{projection} represents the projection property, as is 
summarized in Table II.  We will now show that $X_sX_p$ obeys a similar equation,

\be
   X_sX_p \ket{\Psi} = \ket{\Psi},
\label{superposition}
\ee
which, together with (\ref{projection}), identifies (\ref{state5}) as a generalized Bell 
state and expresses the {\it superposition} property - namely, that \ref{state5} is a 
superposition of the $d$ distinct collapse scenarios.  To prove \Eq{superposition}, 
one can rewrite (\ref{state5}) in terms of the spin and quantum-pointer $X$ bases 
($sX$ and $pX$), whereupon it takes the form
\be
 \ket{\Psi} =  \oord \sum_{k=0}^{d-1} \ket{-k}_{sX} \ket{k}_{pX},
\label{state5F}
\ee   
and \Eq{superposition} follows immediately.  It is instructive to show that 
(\ref{state5}) is in fact the unique solution of the two eigenvalue equations, 
(\ref{projection}) and (\ref{superposition}).  

A comment is in order on the measurement of $X_s$ and $X_p$:  Realizing that 
the readout devices for both the spin and the pointer record standard basis states, 
one must apply ${\cal F}$ (\Eq{Fourier}) to these qudits in order to take their $X$ 
eigenstates into the corresponding $Z$ eigenstates prior to readout.  [Regarding 
the spin, one cannot simply rotate the Stern-Gerlach magnets, because Fourier 
transformation of a qutrit, for example, lies in SU(3) and not in SU(2).]  

Now, suppose that one measures $X_s$ and finds the spin in the eigenstate 
$\ket{k}_{sX}$.   This measurement projects the pointer into the corresponding 
eigenstate of $X_p$, specifically
\be
    \ket{-k}_{pX} = \oord \bigg(\ket{100...0} + \om^{k} \ket{010...0}
    +\om^{2k} \ket{001...0} + ... + \om^{(d-1) k} \ket{000...1} \bigg).
\label{SpecW}
\ee
This is a generalized $W$ state \cite{W state} - it is a superposition of all the pointer 
basis states. Its detection requires the measurements of both $X_s$ and $X_p$.  
It is noteworthy that these are separately complementary to $Z_s^{\dag}$ and $Z_p$, 
respectively, while the products $X_sX_p$ and $Z_s^{\dag} Z_p$ are compatible.  

A final note - the foregoing illustrates what it means to ``observe a 
superposition.''  Ironically, one does not see a superposition as an ambiguity of 
outcomes; one sees it as a definite outcome of a complementary measurement.  
In the above case, the sought-for superposition is an eigenstate of $X_p$.  
A measurement of the spin in the basis $X_s$ projects the quantum pointer 
onto the corresponding eigenstate of $X_p$, whose eigenvalue is perfectly 
correlated with the spin outcome.  An $X_p$ measurement then confirms the
projected quantum pointer state.

\vskip0.7truecm
{\bf Nondestructive Measurements}
\vskip0.4truecm

Up to now, measurements of the observables $N$ and $\Zsdag Z_p$ have been 
constructed from readouts of the individual (local) qubits of the ancilla system, and 
of the spin qudit.  Such local measurements preserve the correlations defining the 
singleness and projection properties, but they destroy the state in the process.  To 
preserve the state, we must reject local information telling us {\it which} qubit is 
excited, or {\it what} the qudit spin value is. In this subsection we show how one
could, in principle, measure correlation properties while rejecting such local
information.

We will describe nondestructive measurements of the 
complete set of commuting observables ($N$, $Z_s^{\dag} Z_p$, and $X_sX_p$).
Since the three are compatible, they can all be measured, in principle, on the same 
state of the same system.  For simplicity in the figures to follow, we shall again
specialize to the case of $d=3$, from which the generalization to arbitrary $d$ 
will be obvious.
\begin{figure}[h!]
\includegraphics[scale=0.35]{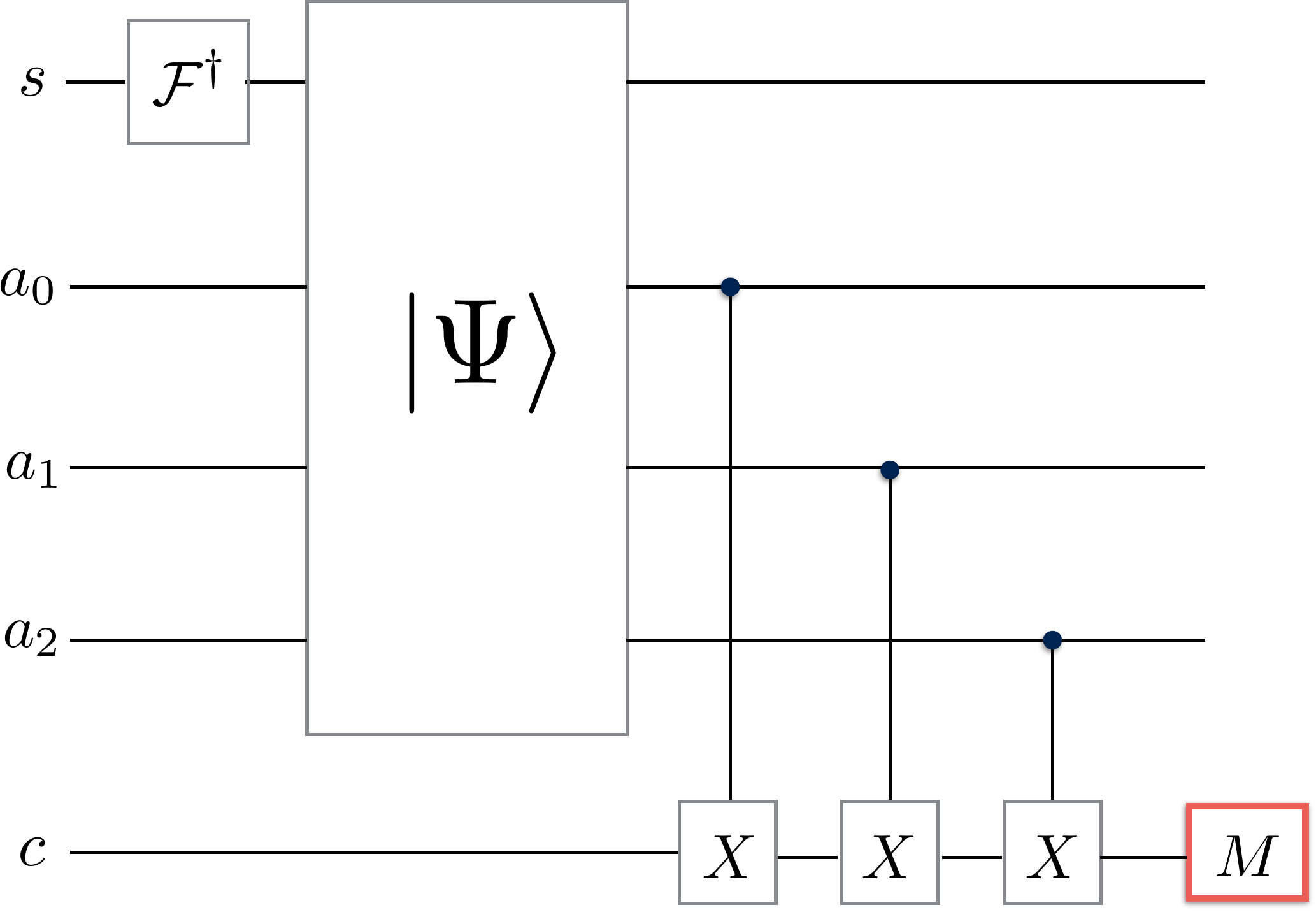}
\caption{\label{fig2} Nondestructive measurement of $N$ on the prepared state 
$\ket{\Psi}$ (\Eq{state4b}):  All five input lines are initialized to 0.  The desired spin 
state (\Eq{state1}) is prepared by (inverse) Fourier transformation, and then entangled 
(large box) with the ancilla qubit system via Eqs. \ref{state2} - \ref{state4b}. The 
measurement of $N$ then commences with the $X$ gates, one for each qubit in 
the ancilla system.}  
\end{figure}

Let us begin with the simplest observable - the ancilla excitation number $N$.  
Figure 2 shows the prepared state $\ket{\Psi}$ of Eq. (\ref{state4b})  (the large box) 
and an added ancilla {\it qutrit} ($c$) to act as a ``counter.''  Each qubit is coupled 
to this counter by a controlled-$X$ gate:  If a qubit is in the 1 state, 
then the counter state index ($k$, the exponent in $\om^k$) is advanced 
by one; if a qubit is in the 0 state, then there is no change.  Hence, with the counter 
index initialized to zero, the readout index will be the total excitation number $N$.  
For a general input state the outcome would be probabilistic, but in the special case 
of the prepared state [Eq. (\ref{state4b}) or more generally (\ref{state5})], it will take 
the value unity with certainty, confirming the singleness property of the pointer.  
\begin{figure}[h!]
\includegraphics[scale=0.48]{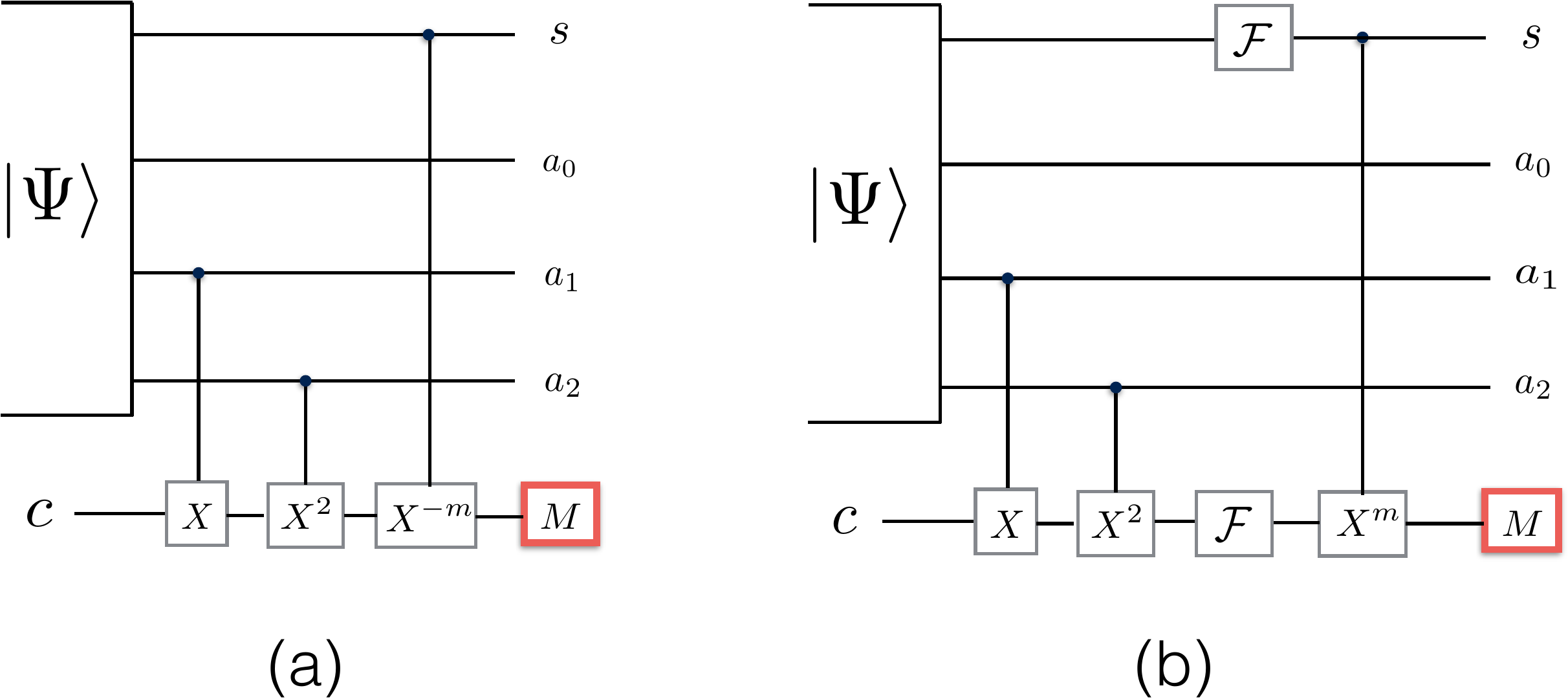}
\caption{\label{fig3} Nondestructive measurements of $Z_s^{\dag} Z_p$ (a) and 
 $X_sX_p$ (b) on the same state $\ket{\Psi}$.}
\end{figure}
%

%
%

Next consider the observable $\Zsdag Z_p$, which represents the projection property.
Figure 3a is a variation on Fig. 2 in which the controlled-$X$ gates are replaced by the 
(qubit-dependent) controlled-$X^l$ gates, where $l$ is the qubit label ($l=0$, 1, or 2, 
with 0 being the identity gate).  Given the singleness property, these gates impress 
the pointer state on the counter qutrit ($c$).  The spin qutrit is then coupled to ($c$) 
by a controlled-$X^{-m}$ gate representing $\Zsdag$, where $m$ labels the 
spin state.  Since the prepared state has $m=l$, this last gate cancels the effect of 
$Z_p$, and the readout exponent will be 0, denoting the eigenvalue 
$\Zsdag Z_p \rightarrow 1 \equiv \om^0$.

Finally, inclusion of the measurement of $X_sX_p$ demonstrates the superposition 
property.  Figure 3b builds upon 3a by including the quantum Fourier transforms
to take $X$ eigenstates into $Z$ eigenstates.  A final controlled-$X^m$ gate couples 
the spin to the counter qutrit ($c$).
Since the prepared state has $m=-l$ in the $X_s$/$X_p$ 
basis, the readout exponent will again be $M=0$.  The upshot of these points is that, 
if we re-initialize the ancilla qutrit ($c$) after each measurement, it is possible in 
principle to measure $N$, $\Zsdag Z_p$, and $X_sX_p$ in any order, on the same 
state of the same system.  

This section comprises a generalization of the Ref. \cite{JL1} treatment to 
arbitrary nonbinary spins.  It is remarkable that the spin-ancilla system is again 
described completely by three commuting observables - two for collapse properties
and one for the superposition property.  Perhaps most remarkable is the emergence
of the quantum pointer through the singleness property - a collapse property resulting 
from the unitary evolution that entangles the spin and the quantum pointer.

\section{CHOICES of the POINTER BASIS}
\label{sec:4}

The previous section shows how the ancilla system provides a choice of pointer 
basis - the basis whose elements are identified with the output of the display pointer.    
Examples discussed were the eigenbases of $Z_p$ and $X_p$.  We note here that 
there is a continuum of such choices, similarly accessible to experiment.

Consider an arbitrary unitary transformation of the standard pointer basis, 
$U_p \ket{k}_p$.  There is a corresponding unitary transformation of the spin basis, 
$V_s \ket{k}_s$, which leaves the entangled state (\ref{state5}) unchanged:  It is straitforward to show that $V$ is the complex conjugate of $U$:
\be
   V_s = U_s^*  \hskip1.0truecm  \Leftrightarrow  \hskip1.0truecm   
   V_sU_p \ket{\Psi} = \ket{\Psi},
\label{invariance}
\ee
These transformed bases are eigenbases of the observables 
\be
   O_p = U_p Z_p U_p^{\dag} \hskip.8truecm \hbox{and} \hskip.8truecm 
    \tilde{O}_s = V_s Z_s V_s^{\dag},
\label{observables O}
\ee
defined respectively in the quantum pointer and spin Hilbert spaces.  We can 
rewrite (\ref{state5}) in the $\tilde{O}$ - $O$ bases (denoted by subscripts) as
\be
     \ket{\Psi} = \oord \sum_{k=0}^{d-1} \ket{k}_{s\tilde{O}} \ket{k}_{pO},
\label{state5U}
\ee
Clearly, this is an eigenstate of $\tilde{O}_s^{\dag} O_p$, which one could verify 
experimentally, in principle, by measuring the $\tilde{O}_s$ component of the spin 
downstream, thus projecting the quantum pointer into the corresponding 
eigenstate of $O_p$.  To measure these observables, according to the discussion
following \Eq{state5F}, one would apply $V_s^{\dag}$ to the spin and $U_p^{\dag}$
to the quantum pointer prior to the readouts.  Incidentally, it is an instructive exercise 
to recover the Fourier transformed representation (\Eq{state5F}) from this prescription 
(see the Appendix if interested).

\subsection{Coherence Requirement}

The necessary condition for basis flexibility and the observation of the superposition 
property (\eg, $\tilde{O}_s O_p  = 1$), within the Stern-Gerlach scenario, is that the 
spatial wave packets must be controlled with sufficient precision to approximate the 
ideal coherent superposition indicated by \Eq{state4a} [and understood in 
(\ref{state4b})].  The challenge has been 
emphasized by Scully \etal~ through the metaphor of Humpty-Dumpty \cite{Scully}.  
The challenge has been met - a coherent Stern-Gerlach interferometer has been 
achieved, but only recently and only for the case of spin-1/2 \cite{SG.interf.19}.  So, 
instead of atomic spins, one might consider alternative platforms, such as photons 
in a Mach-Zender interferometer.  Encouraging recent experimental achievements 
include the entanglement of a photonic qubit with two photonic qutrits \cite{Malik} 
(qutrits made with photon orbital angular momentum), and the demonstration of 
a three-qutrit GHZ state with superconducting transmon circuits \cite{Alba}. 

It is important to emphasize that the coherence requirement for the superposition
property does not apply to collapse properties - neither the projection property 
($Z_s^{\dag} Z_p = 1$) nor the singleness property ($N = 1$).  To show why 
this is the case, consider the worst-case scenario in which there is no coherence 
between the different wave packets when they are brought back to the same 
vicinity at the time $t_4$.  The state of the spin/ancilla system is then described 
by the density matrix, 
\be
   \rho = \third \bigg(  \ket{0}_s \ket{0}_p \bra{0}_p  \bra{0}_s 
    +  \ket{1}_s \ket{1}_p \bra{1}_p \bra{1}_s +  
    \ket{2}_s \ket{2}_p \bra{2}_p \bra{2}_s \bigg),
\label{rho}
\ee
which would replace Eq. (\ref{state4b}).  Clearly the average measured value of 
$X_sX_p$ vanishes,
Tr\big($\rho  X_s X_p\big) = 0$, indicating a uniform probability distribution 
over all possible values, $\omega^k$ ($k = 0,1,...,d-1$), making it impossible 
to observe a superposition.  On the other hand, every term in (\ref{rho}), and 
hence $\rho$ itself, is an eigenstate of both $N$ and $Z_s^{\dag} Z_p$ with 
eigenvalue unity, indicating definite outcomes for both,
Tr\big($\rho  Z_s^{\dag} Z_p\big) = Tr\big(\rho N\big) = 1$.  In the case of 
partial coherence, where nonzero off-diagonal terms such as 
$\ket{0}_s \ket{0}_p \bra{1}_p \bra{1}_s$ contribute to $\rho$, then $X_sX_p$ 
will be biased but still random, while $Z_s^{\dag} Z_p$ and $N$ retain their 
definiteness.   

From the perspective of decoherence theory, the density matrix is derived by
tracing over the imperfectly-controlled spatial degrees of freedom, which are 
playing the role of the environment.  It is not the usual active environment 
responding to the system, but rather a lack of precision in the magnetic field 
configuration and the resulting failure to reverse the separation of paths.  
Nevertheless, it selects the preferred basis associated with ``which path'' 
information, recorded by the display pointer as the eigenvalue of $Z_p$.  The 
reason why $Z_p$ is the ``default'' observable when we lose the flexibility to 
transform the pointer basis is explained in detail below.

\subsection{The Default Basis}

As we said in Sec. III, the $k$th readout device is tuned to record
an eigenstate of the observable $z_k$ of the $k$th qubit.  This observable 
is the (dimensionless) energy of the atom-induced transition $0 \rightarrow 1$.  
The choice to measure $z_k$ is made by designing the qubit-readout 
interaction to depend on this energy.  Since the readout is an irreversible 
decohering process, this interaction selects $z_k$ as the preferred basis of 
an individual qubit.  But this is not yet the preferred basis of the pointer. The 
pointer emerges from the array of all such qubits, each interacting locally
with its own readout device.  Thus there are $2^d$ conceivable readout 
patterns, but because of the singleness property,
only $d$ of these patterns can be realized.  
These represent just the singly-excited states of the ancilla system, which 
are the eigenstates of $Z_p$, so that $Z_p$ is the ``observable of the 
apparatus.''  In summary, the qubit-readout interaction determines the 
preferred basis of individual qubits ($z_k$), while the singleness
property constrains the array to display eigenstates of $Z_p$.

The role of the environment in this determination is restricted to the workings
of individual readout devices.  It prevents the appearance of superpositions of
readings, allowing only a definite 0 or a definite 1.  But it does not assign the 
physical measings ($z_k$) to these outcomes.

\section{POINTER without ANCILLAE}
\label{sec:5}

To identify the quantum pointer in an ordinary apparatus (unaided by
ancillae), consider a Stern-Gerlach setup as in Fig. 1, but with a particle 
detector in place of each ancilla/readout combination.  Clearly 
the detector array forms only the display pointer.   So what is the quantum 
pointer which complements this detector array?  That is, what is it that
entangles with the object of study?

We answer this question by removing the ancilla ket 
from \Eq{state2}.  This leaves an expression identical in form to \Eq{state4b}, 
and we rewrite it as
\bea
    \ket{\Psi(t)} =  & \oort \bigg( \ket{0}_s  \ket{\phi_0(\vec{r},t)} + 
     \ket{1}_s \ket{\phi_1(\vec{r},t)}+ \ket{2}_s \ket{\phi_2(\vec{r},t)} \bigg)    \\
   &  \equiv  \oort \bigg( \ket{0}_s \ket{0,t}_p + 
     \ket{1}_s \ket{1,t}_p + \ket{2}_s \ket{2,t}_p \bigg),
\label{state2new}
\eea
to show the entanglement between the spin and spatial parts of the atomic
state, where the kets $\ket{k,t}_p$ in the second line are shorthand for the 
time dependent wave packets generated by the \Sch~equation for each spin 
component $k$.  These provide valid pointer states during the time interval 
($t_2 < t < t_d$) - that is, after they have become orthogonal but not yet 
arrived at a detector.  Given the singleness property, they are the ``which
path'' states - meaning that if the atom is {\it present} on the $k$th path, 
then it is {\it absent} from all the others.   We define these states to include 
the ``which path'' degree of freedom but not the displacement along the 
path, so that the quantum pointer is a $d$ dimensional system, coming 
into existence at $t_2$ and disappearing at $t_d.$

The observable of the apparatus is the ``which path'' observable $Z_p$ 
whose eigenvalues ($\om^k$) represent the path.   We can define this 
and other observables formally and prove the desired properties by 
starting with projection operators analogous to those of  \cite{JL1}:
\be
     \CP_k = \int_{V(k)} d^3\vec{r}~\ket{\phi_k(\vec{r},t)} \bra{\phi_k(\vec{r},t)}
\label{projectionop}
\ee
projects onto that subvolume of the $k$th path which could be occupied 
during the time interval $t_2 < t < t_d$.   The path occupation number $N$ 
and the pointer observable $Z_p$, analogous to Eqs. (\ref{Nop}) and
(\ref{bigZ}) in Sec. III, are then 
\be
      N = \sum_{k=0}^{d-1} \CP_k     
\label{Nop2}
\ee
\be
      \hbox{and}   \hskip1.0truecm     Z_p = \sum_{k=0}^{d-1} \om^k \CP_k.
\label{bigZ2}
\ee
The state (\ref{state2new}) is an eigenstate of $N$ with eigenvalue unity (the
singleness property), and of $\Zsdag Z_p$ with the same eigenvalue (the
projection property).  So the local measurements of $Z_p$ and $Z_s$ are 
random but perfectly correlated.  The state (\ref{state2new}) is also an 
eigenstate of $X_sX_p$ - a superposition of all possible collapse scenarios,
but it is beyond present technology
to implement $X_p$, since this would require a coherent reconfiguration 
of paths prior to the detection stage.

Finally, each detector ($k$) serves as the readout device for the 
corresponding path ($k$), reading (1,0) for ({\it atom, no atom}) on that path.  
This reading is the eigenvalue of ${\cal P}_k$.   We do not have to engineer an 
interaction in this case - we only have to coordinate each detector with a path.  

%
%


\vskip0.7truecm
{\bf An Exception}
\vskip0.4truecm

An instructive example of a measurement without a {\it quantum} pointer was
pointed out to me by W. K. Wootters.  Recall first that a photon polarization
measurement is analogous to an atomic spin measurement; the role of the
Stern-Gerlach magnetic field is played by a polarizing beam splitter (PBS), 
which entangles the photon's polarization with its path.  Now consider an
unpolarized photon (a completely mixed polarization state), and replace the
PBS with an ordinary (non-polarizing) beam splitter.  We are now measuring
just the photon's path.  What was previously the quantum pointer has now
become the measured system, so a separate pointer is superfluous.  Note 
incidentally that the second link in the von Neumann chain is absent; an 
entangled premeasurement state is not formed, and the path states 
comprise the preferred basis by default.

\section{CONCLUSIONS}
\label{sec:6}

We concluded the introduction by asking three questions.
This summary reviews the answers.

The first major point developed in
this paper is that the pointer is a physical system consisting of two 
distinct parts, so that it can bridge the gap between the object of study 
and an observer (conscious or otherwise).  The ``quantum'' part entangles 
with this object in premeasurement, and then 
the ``display'' part reads the quantum part and records the result.  This 
division of labor is necessary because the display part is an effectively 
classical system - it is entangled with the environment at all times, 
so that it cannot form a pure entangled state with the object as 
conventionally assumed in the von Neumann formalism.  In our first example, 
the ancilla qubits form the {\it quantum} part, while the display part consists of 
the respective readout devices.  We later showed that this model maps onto 
a more realistic (unaided) system, in which particle detectors replace the 
readout devices as the display part, while the path system guiding
the atom forms the quantum part.

The second major point is that unitary quantum theory dictates 
the two ``collapse-like'' properties:  (i) the singleness of outcomes, and (ii)  
the projection property (the post-measurement state update rule), as well 
as the more obviously quantum property, (iii) the superposition property, 
all characterized in Table 2.  All of these are observables of the joint 
object/quantum-pointer system (in its $d^2$-dimensional Hilbert space), and 
as such they are quantum properties.  The three comprise a complete set 
of commuting observables for this system; their three eigenvalues define 
the entangled state of the joint system completely.  The superposition 
property could be observed, in principle, for this object/quantum-pointer 
system, although it cannot apply to the display pointer, which behaves 
classically.

The third major point is that, while 
decoherence is necessary in preventing the appearance of superpositions 
of outcomes, it is not sufficient for determining the preferred basis of the 
pointer.  We showed that the pointer emerges from an apparatus utilizing 
$d$ separate binary  detectors, one for each possible output.  
Decoherence restricts the output of each detector 
to a definite 1 or 0, thus imposing classical 
behavior.  However, the assignment of meaning to the possible outputs 
(providing the post-measurement state updates) must be provided by the 
interaction (or simply the coordination) between the quantum 
and display parts of the pointer, as is spelled out in Subsec. 4.2. 


Let us briefly summarize the implications for foundations: 
First, the naive concept of a single apparatus that can behave either 
quantum mechanically or classically is unrealistic.  Secondly, the 
singleness of outcomes, and the projection property, are quantum 
properties of the entangled premeasurement state.  And finally, 
it is the observer, through the arrangement of the apparatus, 
who determines the preferred basis of the pointer.  The last point
may be the most unconventional of the three, at 
variance with the orthodox theory \cite{Brasil,Zurek.81,Schloss.04}.   
But it depends critically on the first point - the division of the pointer 
into quantum and classical parts.

\bigskip
\centerline{{\bf Acknowledgements}}
\medskip

It is a pleasure to thank Bill Wootters and Brian Odom for enlightening
conversations on the subject of this paper.

\bigskip
\centerline{{\bf Appendix: Recovery of simultaneous Fourier transforms}}
\medskip

Here we show that the simultaneous Fourier transform description of the 
entangled state (\ref{state5}) [based on Eqs. (\ref{Xstates}) - \ref{Fourier})] is 
recovered from the general prescription given in Sec. IV.  The Fourier 
transformed basis, $\ket{k}_{pX} = {\cal F}^{\dag} \ket{k}_p$, corresponds to
\be
   O_p \rightarrow X_p = {\cal F}^{\dag} Z_p {\cal F},
\label{A1}
\ee
so that the general unitary transformations defined in Eqs. \ref{invariance} and 
\ref{observables O} are specialized to 
\be
   U_p = {\cal F}^{\dag}  \hskip 1.5truecm   \hbox{and}   
   \hskip1.5truecm   V_s = U_s^* = {\cal F}.
\label{A2}
\ee
since complex conjugation amounts to inversion of the Fourier transformation.  And 
since this inversion amounts to a sign change of labels, the entangled state 
(\ref{state5}) is rewritten as (\ref{state5F}).  The prescription of Sec. V tells us that
this state is an eigenstate of $\tilde{O}_s^{\dag} O_p$, and that 
\be
   \tilde{O}_s = V_s Z_s V_s^{\dag} = {\cal F} Z_s {\cal F}^{\dag} = X^{\dag}
\label{A3}
\ee
so that
\be
   \tilde{O}_s^{\dag} O_p = X_s X_p,
\label{A4}
\ee
as is needed for consistency with Eq. \ref{superposition}.  This shows that the
Fourier transformed representation of Sec. III is recovered by the general 
prescription of Sec. IV.

\end{document}